\documentclass{article}
\def\abstract#1{\vskip 7mm 
        \begin{center}{\large Abstract}\par \smallskip
                \begin{minipage}[c]{12cm}
                        \small #1
                \end{minipage}
        \end{center}
}
\def\title#1{\begin{center}{\Large\bf #1}\end{center}}
\def\author#1{\vskip 5mm \begin{center}{#1}\end{center}}
\def\address#1{\begin{center}{\it #1}\end{center}}
\makeatletter
\@ifundefined{lesssim}{}{}
\@ifundefined{gtrsim}{}{}
\def\vereq#1#2{\lower3pt\vbox{\baselineskip1.5pt \lineskip1.5pt
\ialign{$\m@th#1\hfill##\hfil$\crcr#2\crcr\sim\crcr}}}
\makeatother
\newcommand{\MR}{{\rm I\!R}}
\begin{document}

\title{%
  General framework \\
  of higher order gauge invariant perturbation
  theory
%  \smallskip \\
%  {\large --- Please use this file to complete your manuscript  ---}
}
\author{%
  Kouji Nakamura\footnote{E-mail:kouchan@th.nao.ac.jp}
}
\address{%
  Department of Astronomical Science, 
  the Graduate University for Advanced Studies, \\
  Osawa, Mitaka, Tokyo 181-8588, Japan.
}

\abstract{
Based on the gauge invariant variables proposed in 
[K.~Nakamura, Prog.~Theor.~Phys.\ {\bf 110} (2003), 723.], 
general framework of the second order gauge invariant perturbation
theory on arbitrary background spacetime is considered.
We derived formulae of the perturbative Einstein tensor of each order,
which have the similar form to the definitions of gauge invariant
variables for arbitrary perturbative fields.
As a result, each order Einstein equation is necessarily given in terms
of gauge invariant variables. 
}

%%%%%%%%%%%%%%%%%%%%%%%%%%%%%%%%%%%%%%%%%%%%%%%%%%%%%%%%%%%%%%%%%%%%%%
%\section{Introduction}
%\label{sec:intro}
%%%%%%%%%%%%%%%%%%%%%%%%%%%%%%%%%%%%%%%%%%%%%%%%%%%%%%%%%%%%%%%%%%%%%%

%****************************************************************

The perturbative approach is one of the popular techniques to
investigate physical systems. 
In particular, this approach is powerful when the construction of
exactly soluble models is difficult.
In general relativity, there are many exact solutions to the Einstein
equation\cite{Exact-solutions} but these are often too idealized to
properly represent natural phenomena.
In this situation, the perturbations around appropriate exact
solutions are useful to investigate realistic situations.
Cosmological perturbation theory\cite{Cosmological-Perturbations} is
now the most commonly used technique, and perturbations of black holes
and stars have been widely studied to obtain descriptions of the
gravitational radiation emitted from them\cite{Black-Hole-perturbations}.

%****************************************************************

In general relativistic perturbations, {\it gauge freedom}, which is
unphysical degree of freedom, arises due to general covariance.
To obtain physically meaningful results, we have to fix these gauge 
freedom or to extract {\it gauge invariant part of perturbations}.
These situations are also seen in the recent investigations of the
oscillatory behavior of gravitating Nambu-Goto
membrane\cite{kouchan-papers,kouchan-flat}, which are concerning 
about the dynamical degree of freedom of gravitating extended objects. 
In these works, it was necessary to distinguish true dynamical
degree of freedom from gauge freedom in perturbations and to develop a
gauge invariant treatment of general relativistic perturbations.

%****************************************************************

In particular,  we discussed comparison of the oscillatory 
behavior of a gravitating string with that of a test string, in
Ref.\cite{kouchan-flat}. 
To do this, we have developed two-parameter general relativistic gauge
invariant perturbation on the Minkowski spacetime, in which one of the
perturbation parameter is the string energy density and the other is
the string oscillation amplitude.
Such multi-parameter perturbations have many other physical situations
to be applied.
The perturbation of spherical stars\cite{Kojima} is one of them, in
which we choose the gravitational field of a spherical star as the
background spacetime for the perturbations, one of the parameters for
the perturbations corresponds to the rotation of the 
star and another is the pulsation amplitude of it.
The effects due to the rotation-pulsation coupling are described in the
higher order.
Even in one-parameter case, it is interest to consider higher
order perturbations.
In particular, Gleiser et al.\cite{Gleiser-etal} reported that the
second order perturbations predict accurate wave form of gravitational
waves. 
Thus, there are many physical situations to which higher order
gauge invariant perturbations with multi-parameter should be applied,
and it is worthwhile to discuss them from general point of view.

%****************************************************************

In this article, we show our treatments of two-parameter higher order
gauge invariant perturbations.
In our treatments, we do not specify the physical meanings of
parameters for perturbation nor background spacetime, though it is
necessary to specify both of them when we apply them to some physical
situations.

%****************************************************************

%%%%%%%%%%%%%%%%%%%%%%%%%%%%%%%%%%%%%%%%%%%%%%%%%%%%%%%%%%%%%%%%%%%%%%
%\section{``Gauge freedom'' of perturbations}
%%%%%%%%%%%%%%%%%%%%%%%%%%%%%%%%%%%%%%%%%%%%%%%%%%%%%%%%%%%%%%%%%%%%%%

%****************************************************************

Though the ``general covariance'' is mathematically formulated in the
concept of spacetime manifolds, it intuitively states that there is no
preferred coordinate system in nature. 
Due to this general covariance, ``gauge freedom'' in general
relativistic perturbations arises.
To explain this ``gauge freedom'', we must remind what we are doing in
perturbation theories.

%****************************************************************

In any perturbation theory, we always treat two spacetime: 
one is the physical spacetime $({\cal M},g_{ab})$ which should be 
described by perturbations; another is the background 
$({\cal M}_{0},{}^{(0)}g_{ab})$ which is prepared for calculations. 
Keeping these two manifolds in our mind, we always write the equation
\begin{equation}
  \label{eq:variable-symbolic-perturbation}
  Q(``p\mbox{''}) = Q_{0}(p) + \delta Q(p).
\end{equation}
where $Q$ is any physical field on ${\cal M}$.
Through Eq.~(\ref{eq:variable-symbolic-perturbation}), we are implicitly
assuming that there exists a map 
${\cal X} : {\cal M}_{0}\rightarrow{\cal M}$ $:$  
$p\in{\cal M}_{0}\mapsto ``p\mbox{''}\in{\cal M}$, which is usually
called a ``gauge choice'' in perturbation
theory\cite{J.M.Stewart-M.Walker11974}.
Namely, $Q(``p\mbox{''})$ in
Eq.~(\ref{eq:variable-symbolic-perturbation}) is a field on 
${\cal M}$ and $``p\mbox{''}\in{\cal M}$.
On the other hand, we should regard that the background value
$Q_{0}(p)$ of $Q(``p\mbox{''})$ and its deviation $\delta Q(p)$ from
$Q_{0}(p)$ in Eq.~(\ref{eq:variable-symbolic-perturbation}) are fields
on ${\cal M}_{0}$ and $p\in{\cal M}_{0}$.
Since Eq.~(\ref{eq:variable-symbolic-perturbation}) is for fields,
it implicitly states that the points $``p\mbox{''}\in{\cal M}$ and
$p\in{\cal M}_{0}$ are same.

%****************************************************************

Note that the gauge choice ${\cal X}$ is not unique when we consider
theories in which general covariance is imposed and this degree of
freedom of ${\cal X}$ is ``gauge freedom'' of perturbations.
If there is a preferred coordinate system on both ${\cal M}_{0}$
and ${\cal M}$, we can choose ${\cal X}$ using this coordinate system.
However, there is no such coordinate system due to general
covariance and we have no guiding principle to choose ${\cal X}$.

%****************************************************************

Based on this understanding of ``gauge'', the gauge
transformation is simply the change of the map ${\cal X}$.
To see this, we only consider the one-parameter perturbation case,
because the essence of discussion for multi-parameter perturbations is
same\cite{kouchan-gauge-inv,Bruni-series}. 
We denote the perturbation parameter by $\epsilon$ and we
consider the $m+1$-dim manifold ${\cal N}={\cal M}\times{\MR}$, 
where $m=\dim{\cal M}$ and $\epsilon\in{\MR}$. 
The background 
${\cal M}_{0}=\left.{\cal N}\right|_{\epsilon=0}$ and the physical 
spacetime 
${\cal M}=\left.{\cal N}\right|_{\MR=\epsilon}=:{\cal M}_{\epsilon}$
are submanifolds embedded in ${\cal N}$. 
On this extended manifold ${\cal N}$, we consider an exponential map
${\cal X}_{\epsilon}$ which is a point identification map from 
${\cal M}_{0}$ to ${\cal M}_{\epsilon}$.
The map ${\cal X}_{\epsilon}$ is a gauge choice discussed above. 
We also consider another gauge choice ${\cal Y}_{\epsilon}$.
The pull-back of each gauge choice maps any field $Q_{\epsilon}$ on 
${\cal M}_{\epsilon}$ to 
${}^{\cal X}Q_{\epsilon} := {\cal X}_{\epsilon}^{*}Q_{\epsilon}$ and 
${}^{\cal Y}Q_{\epsilon} := {\cal Y}_{\epsilon}^{*}Q_{\epsilon}$ on 
${\cal M}_{0}$, respectively.

%****************************************************************

The gauge transformation is induced by the map
$\Phi_{\epsilon}={\cal X}^{-1}_{\epsilon}\circ{\cal Y}_{\epsilon}$,  
\begin{eqnarray}
  {}^{\cal Y}Q_{\epsilon} =
  \left.{\cal Y}^{*}_{\epsilon}Q\right|_{{\cal M}_{0}} 
  =
  \left.\left(
      {\cal Y}^{*}_{\epsilon}
      {\cal X}^{*}_{-\epsilon}
      {\cal X}^{*}_{\epsilon}Q\right)
  \right|_{{\cal M}_{0}} 
  =
  \left.\Phi^{*}_{\epsilon}\left(
      {\cal X}^{*}_{\epsilon}Q\right) 
  \right|_{{\cal M}_{0}}
  =  \Phi^{*}_{\epsilon} {}^{\cal X}Q_{\lambda,\epsilon}.
  \label{eq:Bruni-45} 
\end{eqnarray}
The substitution of expansions 
${}^{\cal X}Q_{\epsilon} = Q_{\epsilon}
+ \epsilon{\pounds}_{u}Q_{\epsilon} 
+ (\epsilon^{2}/2){\pounds}_{u}^{2}Q_{\epsilon} + O(\epsilon^{3})$,
${}^{\cal Y}Q = Q_{\epsilon} + \epsilon{\pounds}_{v}Q_{\epsilon}
+ (\epsilon^{2}/2){\pounds}_{v}^{2}Q_{\epsilon} + O(\epsilon^{3})$,
and
$Q = Q_{0} + \epsilon Q_{1} + (\epsilon^{2}/2) Q_{2} + O(\epsilon^{3})$
into Eq.~(\ref{eq:Bruni-45}) leads each order gauge transforamtion
rules,
\begin{eqnarray}
  {}^{\cal Y}Q_{1} - {}^{\cal X}Q_{1} 
  = {\pounds}_{\xi_{1}}{{}^{\cal X}Q_{0}}, \quad
  {}^{\cal Y}Q_{2} - {}^{\cal X}Q_{2} 
  = 2 {\pounds}_{\xi_{1}}{{}^{\cal X}Q_{1}} +
  \left({\pounds}_{\xi_{2}} + {\pounds}_{\xi_{1}}^{2}\right){{}^{\cal X}Q_{0}},
\end{eqnarray}
where $\xi_{1}^{a}=u^{a}-v^{a}$, $\xi_{2}^{a}=[u,v]^{a}$, and $u^{a}$
($v^{a}$) is the generator of ${\cal X}_{\epsilon}$ 
(${\cal Y}_{\epsilon}$).

%****************************************************************

The gauge transformation 
$\Phi_{\epsilon}={\cal X}^{-1}_{\epsilon}\circ{\cal Y}_{\epsilon}$
also induces the coordinate transformation on ${\cal M}_{\epsilon}$.
A chart $({\cal U},X)$ on ${\cal M}_{0}$ with a gauge choice 
${\cal X}_{\epsilon}$ becomes a chart 
$({\cal X}_{\epsilon}{\cal U},X\circ{\cal X}^{-1}_{\epsilon})$ on 
${\cal M}_{\epsilon}$ ($\{x^{\mu}\}$). 
Another gauge choice ${\cal Y}_{\epsilon}$ induces another chart 
$({\cal Y}{\cal U},X\circ{\cal Y}^{-1})$ on 
${\cal M}_{\epsilon}$ ($\{y^{\mu}\}$).
In the passive point of view, we obtain 
\begin{equation}
  \label{eq:y-x-trans-one-parameter}
  y^{\mu}(q) := x^{\mu}(p) =
  \left(\left(\Phi^{-1}\right)^{*}x^{\mu}\right)(q) 
  = x^{\mu}(q) - \epsilon \xi^{\mu}_{1}(q)
  + \frac{\epsilon^{2}}{2}\left\{-\xi_{2}^{\mu}(q) 
    + \xi^{\nu}_{1}(q)\partial_{\nu}\xi_{1}^{\mu}(q)\right\}
  + O(\epsilon^{3}).
\end{equation}
This includes the additional degree of freedom $\xi^{\mu}_{2}$ and
this does show that the gauge freedom in perturbations is more than
the usual assignment of coordinate labels.

%****************************************************************

Similarly, the gauge transformation rules in two-parameter
perturbations are given by 
\begin{eqnarray}
  \label{eq:variable-expansion-two-parameter}
  {}^{\cal X}Q_{\epsilon,\lambda} := 
  {\cal X}_{\epsilon,\lambda}^{*}Q_{\epsilon,\lambda}
  &=:& \sum^{\infty}_{k,k'=0} \frac{\lambda^{k}\epsilon^{k'}}{k!k'!} 
  \;\;
  \delta^{(k,k')}_{\cal X}Q, \quad \delta^{(0,0)}_{\cal X}Q = Q_{0}, \\
  \label{eq:Bruni-47} 
  \delta^{(p,q)}_{{\cal Y}}Q - \delta^{(p,q)}_{{\cal X}}Q &=& 
  {\pounds}_{\xi_{(p,q)}}Q_{0},
  \quad\quad\quad\quad\quad\quad\quad\quad\quad
  \quad\quad\quad\quad\quad\quad\quad\quad\;
  (p,q) = (0,1), (1,0),\\
  \label{eq:Bruni-49} 
  \delta^{(p,q)}_{\cal Y}Q - \delta^{(p,q)}_{\cal X}Q &=& 
  2 {\pounds}_{\xi_{(\frac{p}{2},\frac{q}{2})}} 
  \delta^{(\frac{p}{2},\frac{q}{2})}_{\cal X}Q 
  + \left\{{\pounds}_{\xi_{(p,q)}} + 
    {\pounds}_{\xi_{(\frac{p}{2},\frac{q}{2})}}^{2}
  \right\} Q_{0},
  \quad\quad\quad (p,q) = (0,2), (2,0),\\
  \label{eq:Bruni-51} 
  \delta^{(1,1)}_{\cal Y}Q - \delta^{(1,1)}_{\cal X}Q &=& 
  {\pounds}_{\xi_{(1,0)}} \delta^{(0,1)}_{\cal X}Q 
  + {\pounds}_{\xi_{(0,1)}} \delta^{(1,0)}_{\cal X}Q 
  \nonumber\\
  && 
  \quad\quad
  + \left\{{\pounds}_{\xi_{(1,1)}} 
    + \frac{1}{2} {\pounds}_{\xi_{(1,0)}}{\pounds}_{\xi_{(0,1)}}
    + \frac{1}{2} {\pounds}_{\xi_{(0,1)}}{\pounds}_{\xi_{(1,0)}}
  \right\} Q_{0}.
\end{eqnarray}

%****************************************************************

%%%%%%%%%%%%%%%%%%%%%%%%%%%%%%%%%%%%%%%%%%%%%%%%%%%%%%%%%%%%%%%%%%%%%%
%\section{Gauge invariant variables}
%%%%%%%%%%%%%%%%%%%%%%%%%%%%%%%%%%%%%%%%%%%%%%%%%%%%%%%%%%%%%%%%%%%%%%

Now, we define gauge invariant variables. 
Our starting point to construct gauge invariant variables is the
assumption which states that we have already known the procedure
to find gauge invariant variables for the linear metric perturbations.
Then, linear metric perturbations ${}^{(1,0)}h_{ab}$
(${}^{(0,1)}h_{ab}$) decomposed as 
\begin{eqnarray}
  {}^{(p,q)}h_{ab} 
  =: {}^{(p,q)}{\cal H}_{ab} 
  + 2 \nabla_{(a}{}^{(p,q)}X_{b)}, 
  \quad
  (p,q) = (1,0), (0,1),
  \label{eq:linear-metric-decomp}
\end{eqnarray}
where ${}^{(p,q)}{\cal H}_{ab}$ (${}^{(p,q)}X_{a}$) is gauge
invariant (variant).
(Henceforth, we omit the gauge index ${\cal X}$.)

%************************************************

This assumption is non-trivial. 
However, once we accept this assumption, we can always find gauge
invariant variables for higher order perturbations\cite{kouchan-gauge-inv}.
In the second order, the metric perturbations are decomposed as 
\begin{eqnarray}
  \label{eq:widehat-H-ab-in-gauge-X-def-2.0}
  {}^{(p,q)}h_{ab}
  &=:&
  {}^{(p,q)}{\cal H}_{ab}
  + 2 {\pounds}_{{}^{(\frac{p}{2},\frac{q}{2})}X} 
  {}^{(\frac{p}{2},\frac{q}{2})}h_{ab}
  + \left(
      {\pounds}_{{}^{(p,q)}X}
    - {\pounds}_{{}^{(\frac{p}{2},\frac{q}{2})}X}^{2} 
  \right)
  {}^{(0)}g_{ab},
  \;
  (p,q) = (2,0), (0,2);
  \\
  {}^{(1,1)}h_{ab}
  &=:&
  {}^{(1,1)}{\cal H}_{ab}
  + {\pounds}_{{}^{(0,1)}X} {}^{(1,0)}h_{ab}
  + {\pounds}_{{}^{(1,0)}X} {}^{(0,1)}h_{ab}
  \nonumber\\
  && \quad\quad\quad
  + \left\{
    {\pounds}_{{}^{(1,1)}X}
    - \frac{1}{2}
      {\pounds}_{{}^{(1,0)}X}
      {\pounds}_{{}^{(0,1)}X}
    - \frac{1}{2}
      {\pounds}_{{}^{(0,1)}X}
      {\pounds}_{{}^{(1,0)}X}
  \right\} {}^{(0)}g_{ab},
\end{eqnarray}
where ${}^{(p,q)}{\cal H}_{ab}$ (${}^{(p,q)}X_{a}$) is gauge
invariant (variant).

%************************************************

Using gauge variant parts ${}^{(p,q)}X_{a}$ of metric perturbations,
gauge invariant variables for an arbitrary field $Q$ other than
metric\cite{kouchan-gauge-inv} are given by  
\begin{eqnarray}
  \label{eq:matter-gauge-inv-def-1.0} 
  \delta^{(p,q)}{\cal Q} &:=&
  \delta^{(p,q)}Q - {\pounds}_{{}^{(p,q)}X}Q_{0},
  \quad\quad\quad\quad\quad\quad\quad\quad
  \quad\quad\quad\quad\quad\quad\quad\quad
  (p,q) = (1,0), (0,1)
  , \\ 
  \label{eq:matter-gauge-inv-def-2.0} 
  \delta^{(p,q)}{\cal Q} &:=&
  \delta^{(p,q)}Q 
  - 2 {\pounds}_{{}^{(\frac{p}{2},\frac{q}{2})}X} 
      \delta^{(\frac{p}{2},\frac{q}{2})}Q 
  - \left\{
    {\pounds}_{{}^{(p,q)}X}
    -{\pounds}_{{}^{(\frac{p}{2},\frac{q}{2})}X}^{2}
  \right\} Q_{0},
  \quad
  (p,q) = (2,0), (0,2)
  , \\
  \label{eq:matter-gauge-inv-def-1.1} 
  \delta^{(1,1)}{\cal Q} &:=&
  \delta^{(1,1)}Q 
  - {\pounds}_{{}^{(1,0)}X} \delta^{(0,1)}Q 
  - {\pounds}_{{}^{(0,1)}X} \delta^{(1,0)}Q 
  \nonumber\\
  && \quad
  - \left\{{\pounds}_{{}^{(1,1)}X} 
    - \frac{1}{2} {\pounds}_{{}^{(1,0)}X} 
                  {\pounds}_{{}^{(0,1)}X} 
    - \frac{1}{2} {\pounds}_{{}^{(0,1)}X} 
                  {\pounds}_{{}^{(1,0)}X}
  \right\} Q_{0}.
\end{eqnarray}

%************************************************

%%%%%%%%%%%%%%%%%%%%%%%%%%%%%%%%%%%%%%%%%%%%%%%%%%%%%%%%%%%%%%%%%%%%%%
%\section{Einstein equations in terms of gauge invariant variables}
%%%%%%%%%%%%%%%%%%%%%%%%%%%%%%%%%%%%%%%%%%%%%%%%%%%%%%%%%%%%%%%%%%%%%%

%************************************************

Next, we show the expression of perturbative Einstein
equations of each order using these gauge invariant variables
defined above.
We consider the pull-back 
${\cal X}_{\epsilon,\lambda}^{*}\tilde{G}_{a}^{\;\;b}$ on 
${\cal M}_{0}$ of the Einstein tensor $\tilde{G}_{a}^{\;\;b}$ on 
${\cal M}$ and expand ${\cal X}_{\epsilon,\lambda}^{*}\tilde{G}_{a}^{\;\;b}$ as
Eq.~(\ref{eq:variable-expansion-two-parameter}).  
In terms of gauge invariant and variant variables of metric
perturbations defined above, the perturbative Einstein tensors are
given by 
\begin{eqnarray}
  \label{eq:linear-Einstein}
  {}^{(p,q)}{G}_{a}^{\;\;b}
  &=&
  {}^{(1)}{\cal G}_{a}^{\;\;b}\left[{}^{(p,q)}{\cal H}_{c}^{\;\;d}\right]
  + {\pounds}_{\displaystyle {}^{(p,q)}_{}X} \;\; G_{a}^{\;\;b},
  \quad\quad\quad\quad
  \mbox{for} \quad (p,q)=(0,1),(1,0) 
  , \\
  \label{eq:second-Einstein-2,0-0,2}
  {}^{(p,q)}G_{a}^{\;\;b}
  &=& 
  {}^{(1)}{\cal G}_{a}^{\;\;b}\left[{}^{(p,q)}{\cal H}_{c}^{\;\;d}\right]
  + {}^{(2)}{\cal G}_{a}^{\;\;b}
     \left[
       {}^{(p,q)}{\cal H}_{c}^{\;\;d},
       {}^{(p,q)}{\cal H}_{c}^{\;\;d}
     \right]
  + 2 {\pounds}_{\displaystyle {}^{(\frac{p}{2},\frac{q}{2})}X} \;\;
    {}^{(\frac{p}{2},\frac{q}{2})}G_{a}^{\;\;b}
  \nonumber\\
  && \quad
  + \left\{
    {\pounds}_{\displaystyle {}^{(p,q)}X}
    - {\pounds}_{\displaystyle {}^{(\frac{p}{2},\frac{q}{2})}X}^{2}
  \right\} G_{a}^{\;\;b}
  \quad\quad\quad\quad
  \mbox{for} \quad (p,q)=(0,2),(2,0) 
  , \\
  \label{eq:second-Einstein-1,1}
  {}^{(1,1)}G_{a}^{\;\;b}
  &=& 
  {}^{(1)}{\cal G}_{a}^{\;\;b}\left[{}^{(1,1)}{\cal H}_{c}^{\;\;d}\right]
  + {}^{(2)}{\cal G}_{a}^{\;\;b}
     \left[
       {}^{(1,0)}{\cal H}_{c}^{\;\;d},
       {}^{(0,1)}{\cal H}_{c}^{\;\;d}
     \right]
  \nonumber\\
  && \quad
  + {\pounds}_{\displaystyle {}^{(1,0)}_{}X} \;\; {}^{(0,1)}{G}_{a}^{\;\;b}
  + {\pounds}_{\displaystyle {}^{(0,1)}_{}X} \;\; {}^{(1,0)}{G}_{a}^{\;\;b}
  \nonumber\\
  && \quad
  + \left\{
    {\pounds}_{\displaystyle {}^{(1,1)}_{}X} 
    - \frac{1}{2} 
    {\pounds}_{\displaystyle {}^{(1,0)}_{}X} 
    {\pounds}_{\displaystyle {}^{(0,1)}_{}X} 
    - \frac{1}{2} 
    {\pounds}_{\displaystyle {}^{(0,1)}_{}X} 
    {\pounds}_{\displaystyle {}^{(1,0)}_{}X} 
  \right\} G_{a}^{\;\;b}
  ,
  \\
  {}^{(1)}{\cal G}_{a}^{\;\;b}\left[A_{c}^{\;\;d}\right]
  &:=&
  - 2 \nabla_{[a}A_{d]}^{\;\;\;bd} - A^{cb} R_{ac}
  + \frac{1}{2} \delta_{a}^{\;\;b}
  \left(
    2 \nabla_{[e}A_{d]}^{\;\;\;ed} + R_{ed} A^{ed}
  \right)
  , \\
  {}^{(2)}{\cal G}_{a}^{\;\;b}
     \left[
       A_{c}^{\;\;d},
       B_{c}^{\;\;d}
     \right]
     &:=& 
    2 R_{ad} B_{c}^{\;\;(b}A^{d)c}
  + 2 A_{[a}^{\;\;\;de} B_{d]\;\;e}^{\;\;\;b}
  + 2 B_{[a}^{\;\;\;de} A_{d]\;\;e}^{\;\;\;b}
  \nonumber\\
  && \quad
  + 2 A_{e}^{\;\;d} \nabla_{[a}B_{d]}^{\;\;\;be}
  + 2 B_{e}^{\;\;d} \nabla_{[a}A_{d]}^{\;\;\;be}
  + 2 A_{c}^{\;\;b} \nabla_{[a}B_{d]}^{\;\;\;cd}
  + 2 B_{c}^{\;\;b} \nabla_{[a}A_{d]}^{\;\;\;cd}
  \nonumber\\
  && \quad
  - \frac{1}{2} \delta_{a}^{\;\;b}
  \left(
      2 R_{de} B_{f}^{\;\;(d} A^{e)f} 
    + 2 A_{[f}^{\;\;\;de} B_{d]\;\;e}^{\;\;\;f}
    + 2 B_{[f}^{\;\;\;de} A_{d]\;\;e}^{\;\;\;f}
    + 2 A_{e}^{\;\;d} \nabla_{[f}B_{d]}^{\;\;\;fe}
  \right.
  \nonumber\\
  && \quad\quad\quad\quad
  \left.
    + 2 B_{e}^{\;\;d} \nabla_{[f}A_{d]}^{\;\;\;fe}
    + 2 A^{fe} \nabla_{[f}B_{d]e}^{\;\;\;\;\;d}
    + 2 B^{fe} \nabla_{[f}A_{d]e}^{\;\;\;\;\;d}
  \right),
\end{eqnarray}
where $A_{abc}:=\nabla_{(a}A_{b)c}-(1/2)\nabla_{c}A_{ab}$ and
$B_{abc}$ and ${}^{(p,q)}{\cal H}_{abc}$ follow the same definition.
We note that ${}^{(1)}{\cal G}_{a}^{\;\;b}\left[*\right]$ and 
${}^{(2)}{\cal G}_{a}^{\;\;b}\left[*,*\right]$ are gauge invariant
parts of the perturbative Einstein tensors and each expression 
(\ref{eq:linear-Einstein})-(\ref{eq:second-Einstein-2,0-0,2})
has the similar form to
Eqs.~(\ref{eq:matter-gauge-inv-def-1.0})-(\ref{eq:matter-gauge-inv-def-1.1}),
respectively.

%************************************************

Next, to consider the Einstein equation, we expand the energy
momentum tensor as Eq.~(\ref{eq:variable-expansion-two-parameter})
and we impose the perturbed Einstein equation of each order
${}^{(p,q)}G_{a}^{\;\;b} = 8\pi G \;\; {}^{(p,q)}T_{a}^{\;\;b}$.
We defining the each order gauge invariant variable 
${}^{(p,q)}{\cal T}_{a}^{\;\;b}$ for the perturbative energy momentum
tensor by 
Eqs.~(\ref{eq:matter-gauge-inv-def-1.0})-(\ref{eq:matter-gauge-inv-def-1.1}).
Then, perturbative Einstein equation of each order is given by 
\begin{eqnarray}
  \label{eq:each-order-perturbations}
  8\pi G \;\; {}^{(p,q)}{\cal T}_{a}^{\;\;b}
  &=&
  {}^{(1)}{\cal G}_{a}^{\;\;b}\left[{}^{(p,q)}{\cal H}_{c}^{\;\;d}\right],
  \quad\quad\quad\quad\quad\quad\quad\quad\quad\quad\quad\quad\quad\;
  \mbox{for} \quad (p,q)=(0,1),(1,0) 
  , \\
  8\pi G \;\; {}^{(p,q)}{\cal T}_{a}^{\;\;b}
  &=&
  {}^{(1)}{\cal G}_{a}^{\;\;b}\left[{}^{(p,q)}{\cal H}_{c}^{\;\;d}\right]
  + {}^{(2)}{\cal G}_{a}^{\;\;b}
     \left[
       {}^{(p,q)}{\cal H}_{c}^{\;\;d},
       {}^{(p,q)}{\cal H}_{c}^{\;\;d}
     \right],
  \quad
  \mbox{for} \quad (p,q)=(0,2),(2,0), 
  \\
  8\pi G \;\; {}^{(1,1)}{\cal T}_{a}^{\;\;b} &=&
  {}^{(1)}{\cal G}_{a}^{\;\;b}\left[{}^{(1,1)}{\cal H}_{c}^{\;\;d}\right]
  + {}^{(2)}{\cal G}_{a}^{\;\;b}
     \left[
       {}^{(1,0)}{\cal H}_{c}^{\;\;d},
       {}^{(0,1)}{\cal H}_{c}^{\;\;d}
     \right]
  .
\end{eqnarray}
Thus, order by order Einstein equations are necessarily given in
terms of gauge invariant variables only.

%************************************************

%%%%%%%%%%%%%%%%%%%%%%%%%%%%%%%%%%%%%%%%%%%%%%%%%%%%%%%%%%%%%%%%%%%%%%
%\section{Summary}
%%%%%%%%%%%%%%%%%%%%%%%%%%%%%%%%%%%%%%%%%%%%%%%%%%%%%%%%%%%%%%%%%%%%%%

%************************************************

In summary, we showed the general framework of higher order gauge
invariant perturbations of Einstein equation.
We have confirm two facts 
First, {\it if the linear order gauge invariant perturbation
  theory is well established, its extension to higher order and
  multi-parameter perturbation is straightforward}. 
Second, {\it perturbative Einstein equations of each order are
  necessarily given in gauge invariant form}.
The second result is trivial because any equation can be written
in the form that the right hand side is equal to ``zero'' in any
gauge. This ``zero'' is gauge invariant.
Then the left hand side of this equation should be gauge invariant.
In this sense, the above second result is trivial.
However, we have to note that this trivial result implies that our
framework is mathematically correct at this level.

%************************************************

Further, we also note that in our framework, we do not specify
anything about the background spacetime and physical meaning of
the parameters for the perturbations.
Our framework is based only on general covariance.
Hence this framework is applicable to any theory in which general
covariance is imposed and it has very many applications.
Though this framework is not complete, yet, we are planning to apply
this to some physical problems. 
We leave these applications as our future works.

%************************************************

The author would like to thank to Prof. M.~Omote (Keio Univ.) and
Prof. S.~Miyama (NAOJ) for their continuous encouragement.

%************************************************


\begin{thebibliography}{99}
\bibitem{Exact-solutions} H.~Stephani, D.~Kramer, M.~MacCallum,
  C.~Hoenselaers and E.~Herlt, {\it Exact solutions of Einstein's
    Field Equations}, (Second Edition, Cambridge: Cambridge University
  Press, 2003).
\bibitem{Cosmological-Perturbations}
%\bibitem{Bardeen1980}
  J.~M.~Bardeen, Phys.~Rev.\ D\ {\bf 22} (1980), 1882
  ;
  %.
%\bibitem{Kodama-Sasaki}
  H.~Kodama and M.~Sasaki, Prog.~Theor.~Phys.~Suppl.\ No.78 (1984), 1
  ;
  %.
%\bibitem{Mukhanov-Feldman-Brandenberger}
  V.~F.~Mukhanov, H.~A.~Feildman, and R.~H.~Brandenberger,
  Phys.~Rep.\ {\bf 215} (1992), 203. 
\bibitem{Black-Hole-perturbations}
  S.~Chandrasekhar, {\it The mathematical theory of black holes} 
  (Oxford: Clarendon Press, 1983)
  ;
  R.~J.~Gleiser, C.~O.~Nicasio, R.~H.~Price and J.~Pullin,
  Phys.~Rep.\ {\bf 325} (2000), 41
  ;
  K.~D.~Kokkotas and B.~G.~Schmidt, Living Rev. Relativity\ 
  {\bf 2} (1999), 2. 
\bibitem{kouchan-papers}
  K.~Nakamura, A.~Ishibashi and H.~Ishihara, Phys.~Rev.\ D{\bf 62} (2000),
  101502(R);
  K.~Nakamura and H.~Ishihara, Phys.~Rev.\ D\ {\bf 63} (2001), 127501;
%\bibitem{kouchan-loop-initial}
  K.~Nakamura, Class.~Quantum~Grav.\ {\bf 19} (2002), 783;
%\bibitem{kouchan-cylindrical}
  K.~Nakamura, Phys.~Rev.\ D\ {\bf 66} (2002), 084005.
\bibitem{kouchan-flat}
  K.~Nakamura, Prog.~Theor.~Phys. {\bf 110}, (2003), 201.
\bibitem{kouchan-gauge-inv}
  K.~Nakamura, Prog.~Theor.~Phys. {\bf 110}, (2003), 723.
\bibitem{Kojima}
  Y.~Kojima, Prog.~Theor.~Phys.~Suppl.\ No.128 (1997), 251. 
\bibitem{Gleiser-etal}
  R.J.~Gleiser, C.O.~Nicasio, R.H.~Price, and J.~Pullin,
  Phys.~Rev.~Lett.\ {\bf 77} (1996), 4483. 
\bibitem{J.M.Stewart-M.Walker11974}
  J.~M.~Stewart and M.~Walker, Proc.~R.~Soc.~London\ A {\bf 341}
  (1974), 49;
  %. \\
  J.~M.~Stewart, Class.~Quantum~Grav.\ {\bf 7} (1990), 1169;
  %. \\
  J.~M.~Stewart, {\it Advanced General Relativity} (Cambridge
  University Press, Cambridge, 1991).
\bibitem{Bruni-series}
  M.~Bruni, S.~Matarrese, S.~Mollerach and S.~Sonego,
  Class. Quantum Grav.\ {\bf 14}, (1997), 2585;
  M.~Brun and S.~Sonego, Class.~Quantum~Grav.\ {\bf 16} (1999), L29;
  M.~Bruni, L.~Gualtieri, and C.~F.~Sopuerta, Class.~Quantum~Grav.\
  {\bf 20}, (2003), 535; C.F.~Sopuerta, M.~Bruni, L.~Gualtieri,
%  ``Nonlinear N parameter space-time perturbations: gauge
%  transformations through the Baker-Campbell-Hausdorff formula'',
  arXiv:gr-qc/0306027. 
\end{thebibliography}
\end{document}